\documentclass[aps,twocolumn,pre,superscriptaddress]{revtex4mod}
\usepackage[utf8]{inputenc}
\usepackage{graphicx}
\usepackage{amsfonts}
\usepackage{amsmath}
\usepackage{amssymb}

\newcommand{\Refcite}[1]{Ref.~\cite{#1}}
\newcommand{\Refscite}[1]{Refs.~\cite{#1}}
\newcommand{\Eqref}[1]{Eq.~\eqref{#1}}
\newcommand{\Eqsref}[1]{Eqs.~\eqref{#1}}
\newcommand{\Secref}[1]{Sec.~\ref{#1}}
\newcommand{\Figref}[1]{Fig.~\ref{#1}}

\newcommand{\GENERIC}{\textsc{generic}}
\newcommand{\BLOB}{\textsc{blob}}

\newcommand{\latin}[1]{{\itshape #1}}
\newcommand{\abinitio}{\latin{ab initio}}

\newcommand{\defacto}{\latin{de facto}}

\newcommand{\etal}{\latin{et al.\/}}
\newcommand{\etc}{\latin{et\,c}}
\newcommand{\ie}{\latin{i.$\,$e.}}

\newcommand{\viceversa}{\latin{vice versa}}

\newcommand{\kB}{k_{\mathrm{B}}}
\newcommand{\kT}{\kB T}

\newcommand{\myvec}[1]{{\mathbf #1}}
\newcommand{\rvec}{\myvec{r}}
\newcommand{\Fvec}{\myvec{F}}
\newcommand{\evec}{\myvec{e}}
\newcommand{\wc}{\omega^c}

\begin{document}

\title{Perspective: Dissipative Particle Dynamics}

\author{Pep  Espa\~{n}ol}

\affiliation{Dept.\ F\'{\i}sica Fundamental, Universidad Nacional
  de Educaci\'on a Distancia, Aptdo.\ 60141 E-28080, Madrid, Spain}

\author{Patrick B. Warren}

\affiliation{Unilever R\&D Port Sunlight,
Quarry Road East, Bebington, Wirral, CH63 3JW, UK.}

\date{December 13, 2016}

\begin{abstract}
Dissipative particle dynamics (DPD) belongs to a class of models and
computational algorithms developed to address mesoscale problems in
complex fluids and soft matter in general.  It is based on the notion
of particles that represent coarse-grained portions of the system
under study and allow, therefore, to reach time and length scales that
would be otherwise unreachable from microscopic simulations.  The
method has been conceptually refined since its introduction almost
twenty five years ago.  This perspective surveys the major conceptual
improvements in the original DPD model, along with its microscopic
foundation, and discusses outstanding challenges in the field. We
summarize some recent advances and suggests avenues for future
developments.
\end{abstract}

\maketitle

\section{Introduction}
The behaviour of complex fluids and soft matter in general is
characterized by the presence of a large range of different time and
space scales.  Any attempt to resolve \emph{simultaneously} several
time scales in a \emph{single} simulation scheme is confronted by the
problem of taking a prohibitively large number of sufficiently small
time steps.  Typically one proceeds hierarchically
\cite{Berendsen2007}, by devising models and algorithms appropriate to
the length and time scales one is interested in.  Leaving aside
quantum effects negligible for soft matter, at the bottom of the
hierarchy we have Hamilton's equations, with accurate albeit
approximate potential energy functions, which are solved numerically
with molecular dynamics (MD).  Nowadays some research teams can
simulate billions of particles for hundreds of nanoseconds
\cite{Heinecke2015}.  This opens up the possibility to study very
detailed, highly realistic molecular models that capture essentially
all the microscopic details of the system.  This is, of course, not
enough in many situations encountered in soft matter and life sciences
\cite{Shillcock2008}.  One can always think of a problem well beyond
computational capabilities: from the folding of large proteins, to the
replication of DNA, or the simulation of an eukariotic cell, or the
simulation of a mammal, including its brain.  While we are still very
far from even well-posing some of these problems, it is obvious that
science is pushing strongly towards more and more complex systems.

Instead of using atoms moving with Hamilton's equations to describe
matter, one can take a continuum approach in which fields take the
role of the basic variables. Navier-Stokes-Fourier hydrodynamics, or
elasticity, or many of the different continuum theories for complex
fluid systems are examples of this approach \cite{Ottinger2005}.
These continuum theories are, in fact, coarse-grained versions of the
atomic system that rely on two key related concepts: 1) the continuum
limit---\ie\ a ``point'' of space on which the field is defined is, in
fact, a volume element containing a large number of atoms
\cite{Batchelor1967}, and 2) the local equilibrium
assumption---\ie\ these volumes are large enough to reproduce the
thermodynamic behaviour of the whole system \cite{deGrootMazur1964}.
The quantities from one volume element to its neighbour are assumed to
change little and this allows the powerful machinery of partial
differential equations to describe mathematically the system at the
largest scales, allowing even to find analytical solutions for many
situations.  Nevertheless, the continuum equations are usually
non-linear and analytical solutions are not always possible.  One
resorts then to numerical methods to solve the equations.
Computational fluid dynamics (CFD) has evolved into a sophisticated
field in numerical analysis with a solid mathematical foundation.

The length scales that can be addressed by continuum theories range
from microns to parsecs. Remarkably, the same equations (with the same
thermodynamics and transport coefficients) can be used at very
different scales. Many of the interesting phenomena that occur in
complex fluids occur at the \emph{mesoscale}.  The mesoscale can be
roughly defined as the spatio-temporal scales ranging from
10--$10^4$\,nm and 1--$10^6$\,ns.  These scales require a number of
atoms that make the simulation with MD readily unfeasible.  On the
other hand, it was shown in the early days of computer simulations by
Alder and Wainwright \cite{Alder1970} that hydrodynamics is valid at
surprisingly small scales.  Therefore, there is a chance to use
continuum theory down to the mesoscale.  However, at these short
length scales the molecular discreteness of the fluid starts to
manifest itself.  For example, a colloidal particle of submicron size
experiences Brownian motion which is negligible for macroscopic bodies
like submarine ships.  In order to address these small scales one
needs to equip field theories like hydrodynamics with fluctuating
terms, as pioneered by Landau and Lifshitz \cite{Landau1959}.  The
resulting equations of fluctuating hydrodynamics also receive the name
of Landau-Lifshitz-Navier-Stokes (LLNS) equation.  There is much
effort in the physics/mathematical communities to formulate numerical
algorithms with the standards of usual CFD for the solution of
stochastic partial differential equations modeling complex fluids at
mesoscales \cite{Naji2009, Uma2011, Shang2012, Donev2010, Oliver2013,
  Donev2014, Donev2014a, Plunkett2014, DeCorato2016}.

While the use of fluctuating hydrodynamics may be appropriate at the
mesoscale, there are many systems for which a continuum hydrodynamic
description is not applicable (or it is simply unknown).  Proteins,
membranes, assembled objects, polymer systems \etc.  may require
unaccessible computational resources to be addressed with full
microscopic detail but a continuum theory may not exist.  In these
mesoscale situations, the strategy to retain some chemical specificity
is to use \emph{coarse-grained} descriptions in which groups of atoms
are treated as a unit \cite{Voth2009}.  While the details of how to do
this are very system specific, and an area of intense active research
(see reviews in \Refscite{Noid2013, Brini2013, Lopez2014}), it is good
to know that there is a well defined and sounded procedure for the
construction of coarse-grained descriptions \cite{Green1952,
  Zwanzig1961} that is known under the names of non-equilibrium
statistical mechanics, Mori-Zwanzig theory, or the theory of
coarse-graining \cite{Grabert1982, Zubarev1996, Espanol2004Chapter,
  Ottinger2005}. Simulating everything, everywhere, with molecular
detail can be not only very expensive but also unnecessary.  In
particular, water is very expensive to simulate and sometimes its
effect is just to propagate hydrodynamics.  Hence there is an impetus
to develop at least coarse-grained \emph{solvent} models, but retain
enough \emph{solute} molecular detail to render chemical specificity.

At the end of the 20th century the simulation of the mesoscale was
attacked from a computational point of view with a physicist
intuitive, quick and dirty, approach.  Dissipative particle dynamics
(DPD) was one of the products, among others \cite{Malevanets1999,
  Succi2001, KHbook09, Dunweg2009, Gompper2009, Donev2009}, of this
approach.  DPD is a point particle minimal model constructed to
address the simulation of fluid and complex systems at the mesoscale,
when hydrodynamics and thermal fluctuations play a role.  The
popularity of the model stems from its algorithmic simplicity and its
enormous versatility.  Just by varying at will the conservative forces
between the dissipative particles one can readily model complex fluids
like polymers, colloids, amphiphiles and surfactants, membranes,
vesicles, phase separating fluids, \etc.  Due to its simple
formulation in terms of symmetry principles (Galilean, translational,
and rotational invariances) it is a very useful tool to explore
generic or universal features (scaling laws, for example) of systems
that do not depend on molecular specificity but only on these general
principles.  However, detailed information highly relevant for
industrial and technological processes requires the inclusion of
chemical detail in order to go beyond qualitative descriptions.

DPD,  as  originally  formulated,   does  not  include  this  chemical
specificity.  This is  not a drawback of  DPD per se, as  the model is
regarded   as   a  coarse-grained   version   of   the  system.    Any
coarse-graining process  eliminates details  from the  description and
keeps only the relevant ones associated  to the length and time scales
of the  level of description under  scrutiny.  However, as it  will be
apparent,  the original  DPD  model  could be  regarded  as being  too
simplistic and  one can  formulate models  that capture  more accurate
information of the system with comparable computational efficiency.

Since its initial introduction, the question ``What do the dissipative
particles represent?'' has lingered in the literature, with
intuitively appealing but certainly vague answers like ``groups of
atoms moving coherently''. In the present Perspective we aim at
answering this question by reviewing the efforts that have been taken
in this direction.  We offer a necessarily brief overview of
applications, and discuss some open questions and unsolved problems,
both of fundamental and applied nature. Since the initial formulation
of the DPD model a number of excellent reviews \cite{Warren1998,
  Espanol2004Chapter, Pivkin2010, Moeendarbary2010a, Guigas2011,
  Lu2013, Ghoufi2013, Liu2014}, and dedicated workshops
\cite{CECAM2008, Mousseau2014}, have kept the pace of the
developments.  We hope that the present perspective complements these
reviews with a balanced view about the more recent advances in the
field.  We also provide a route map through the different DPD variant
models and their underlying motivation. In this doing, we hope to
highlight a unifying conceptual view for the DPD model and its
connection with the microscopic and continuum levels of description.

This Perspective  is organized  as follows.  In \Secref{Sec:DPD} we
consider the original  DPD model with its virtues  and limitations. In
\Secref{Sec:MDPD} we  review models  that have  been formulated  in
order to  avoid the limitations of  the original DPD model.   The SDPD
model,  which is  the culmination  of the  previous models  that links
directly to  the macroscopic  level of description  (Navier-Stokes) is
considered in  \Secref{Sec:topdown}.  The microscopic  foundation of
the DPD  model is  presented in  \Secref{Sec:bottomup}.  Finally, we
present some selected applications  in \Secref{Sec:applications} and
conclude in \Secref{Sec:conclusions}.

\section{The original DPD model}\label{Sec:DPD}
The original DPD model was introduced by Hoogerbrugge and Koelman
\cite{Hoogerbrugge1992}, and was formulated by the present authors as
a proper statistical mechanics model shortly after
\cite{Espanol1995epl}.  The DPD model consists on a set of point
particles that move off-lattice interacting with each other with three
types of forces: a conservative force deriving from a potential, a
dissipative force that tries to reduce radial velocity differences
between the particles, and a further stochastic force directed along
the line joining the center of the particles.  The last two forces can
be termed as a ``pair-wise Brownian dashpot'' which, as opposed to
ordinary Langevin or Brownian dynamics, is momentum conserving. The
Brownian dashpot is a minimal model for representing viscous forces
and thermal noise between the ``groups of atoms'' represented by the
dissipative particles.  Because of momentum conservation the behaviour
of the system is hydrodynamic at sufficiently large scales
\cite{Espanol1995pre, Marsh1997, BJM15}.

\begin{figure}[t]
    \centering
    \includegraphics[scale=0.15]{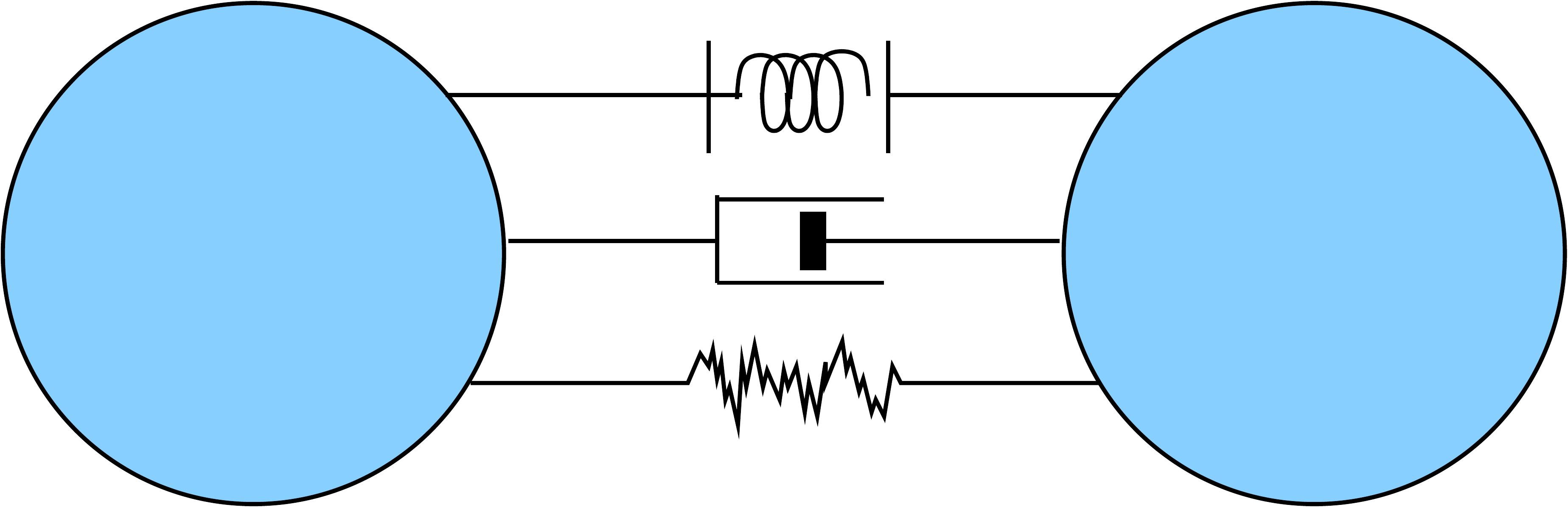}
    \caption{Dissipative   particles   interact   pair-wise   with   a
      conservative linear repulsive force, and a Brownian dashpot made
      of a friction  force that reduces the  relative velocity between
      the particles and  a stochastic force that gives  kicks of equal
      size  and opposite  directions  to the  particles. These  forces
      vanishes beyond a cutoff radius $r_c$.}
    \label{Fig.Dashpot}
\end{figure}

The stochastic  differential equations  of motion for  the dissipative
particles are \cite{Espanol1995epl}
\begin{align}
  \dot{\bf r}_i=& {\bf v}_i
\nonumber\\
m_i\dot{\bf v}_i=& -\frac{\partial V}{\partial {\bf r}_i}
-\sum_j \gamma \omega^D(r_{ij}) ({\bf v}_{ij} \cdot {\bf e}_{ij}){\bf e}_{ij}
\nonumber\\
&+\sum_j \sigma \omega^R(r_{ij}) \frac{dW_{ij}}{dt}{\bf e}_{ij}
  \label{dpd}
\end{align}
where $r_{ij}=|{\bf r}_i-{\bf r}_j|$ is the distance between particles
$i,j$, ${\bf v}_{ij}={\bf v}_{i}-{\bf v}_{j}$ is the relative velocity
and  ${\bf e}_{ij}={\bf  r}_{ij}/r_{ij}$  is the  unit vector  joining
particles $i$ and  $j$.  $dW_{ij}$ is an independent  increment of the
Wiener   process.   In   \Eqref{dpd},  $\gamma$   is  a   friction
coefficient  and  $\omega^D(r_{ij}),\omega^R(r_{ij})$ are  bell-shaped
functions  with   a  finite   support  that  render   the  dissipative
interactions local.   Validity of the  fluctuation-dissipation theorem
requires \cite{Espanol1995epl}  $\sigma$ and $\gamma$ to  be linked by
the    relation     $\sigma^2=2    \gamma    \kT$     and    also
$\omega^D(r_{ij})=[\omega^R(r_{ij})]^2$.   Here  $\kB$ is  Boltzmann's
constant  and  $T$  is  the  system temperature.   As  a  result,  the
stationary probability distribution  of the DPD model is  given by the
Gibbs canonical ensemble
\begin{align}
  \rho(\{{\bf r},{\bf v}\}) &
  =\frac{1}{Z}\exp\left\{-\beta\sum_i^Nm_i\frac{{\bf v}_i^2}{2}
  -\beta V(\{{\bf r}\})\right\}
\label{gibbs}
\end{align}
The potential energy $V(\{{\bf r}\})$ is a suitable function of the
positions of the dissipative particles that is translationally and
rotationally invariant in order to ensure linear and angular momentum
conservation.  In the original formulation the form of the potential
function was taken of the simplest possible form
\begin{align}
V(\{{\bf r}\})&=\frac{1}{2}\sum_{ij}a_{ij}(1-r_{ij}/r_c)^2  
\label{V}
\end{align}
where $a_{ij}$ is a particle interaction constant and $r_c$ is a
cutoff radius.  This potential produces a linear force with the form
of a Mexican hat function of finite range.  Without any other
guidance, the weight function $\omega^R(r)$ in the dissipative and
random forces is given by the same linear functional form.  Complex
fluids can be modeled through mesostructures constructed by adding
additional interactions (springs and/or attractive or repulsive
potentials between certain particles) to the particles. Groot and
Warren \cite{Groot1997} offered a practical route to select the
parameters in a DPD simulation by matching compressibility and
solubility parameters of the model to real systems.

The soft nature of the weight functions in DPD allows for large time
steps, as compared with MD that needs to deal with steep repulsive
potentials.  However too large time steps lead to numerical errors
that depend strongly on the numerical algorithm used.  The area of
numerical integrators for the stochastic differential equations of DPD
has received attention during the years with increasingly
sophisticated methods.  Starting from the velocity Verlet
implementation of \Refcite{Groot1997} and the self-consistent
reversible scheme of Pagonabarraga and Hagen \cite{Pagonabarraga1998},
the field has evolved towards splitting schemes \cite{Shardlow2003,
  Nikunen2003, Defabritiis2006, Serrano2006, Thalmann2007,
  Chaudhri2010}.  A Shardlow \cite{Shardlow2003} scheme has been
recommended after comparison between different integrators
\cite{Nikunen2003}, but there are also other recent more efficient
proposals \cite{Litvinov2010, Leimkuhler2015, Farago2016}.

Because of momentum conservation, the original DPD model in
\Eqsref{dpd}--\eqref{V} can be regarded as a (toy) model for the
simulation of fluctuating hydrodynamics of a simple fluid.  As a model
for a Newtonian fluid at mesoscales the DPD model has been used for
the simulation of hydrodynamics flows in several situations
\cite{JBK+99, Keaveny2005, Chen2006a, VandeMeent2008, Tiwari2008,
  Steiner2009, Filipovic2011}.  It should be obvious, though, that the
fact that DPD conserves momentum does not makes it the preferred
method for solving hydrodynamics. MD is also momentum conserving and
can be used to solve hydrodynamics; for a recent review see Kadau
\etal\ \cite{Kadau2010}.  However, in terms of computational efficiency
hydrodynamic problems are best addressed with CFD methods with,
perhaps, inclusion of thermal fluctuations.

In addition, the original DPD model suffers from several limitations
that downgrade its utility as a LLNS solver.  The first one is the
thermodynamic behaviour of the model.  Taken as a particle method, the
DPD model has an equation of state that is fixed by the conservative
interactions.  The linear conservative forces of the original DPD
model produce an unrealistic equation of state that is quadratic in
the density \cite{Groot1997}.  The quadratic equation of state in DPD
seems to be a general property of soft sphere fluids at high overlap
density.  A well-known exemplar is the Gaussian core model
\cite{Louis2000}.  These systems have been termed \emph{mean-field}
fluids and this includes the linear DPD potential in \Eqref{V}.  Many
thermodynamic properties for the linear DPD potential can be obtained
by using standard liquid state theory and it has been our experience
that the HNC integral equation closure works exceptionally well in
describing the behaviour of DPD in the density regime of interest
\cite{WVA+13, Warren2014, FvM+16}.  Note that while it is possible to
fit the compressibility (related to second derivatives of the free
energy) to that of water, for example, the pressure (related to first
derivatives) turns out to be unrealistic.  The conservative forces of
the original model are not flexible enough to specify the
thermodynamic behaviour as an input of the simulation code
\cite{Trofimov2002}.

A second limitation is due to too simplistic friction forces.  The
central friction force in \Eqref{dpd} implies that when a dissipative
particle passes a second, reference particle, it will not exert any
force on the reference particle unless there is a radial component to
the velocity \cite{Espanol1997epl, Espanol1998pre}.  Nevertheless, on
simple physical grounds one would expect that the passing dissipative
particle would drag in some way the reference particle due to shear
forces.  Of course, if many DPD particles are involved simultaneously
in between the two particles, this will result in an effective
drag. The same is true for a purely conservative molecular dynamics
simulation.  It would be nice, though, to have this effect captured
directly in terms of modified friction forces in a way that a smaller
number of particles need to be used to reproduce large scale
hydrodynamics.  Note that the viscosity of the DPD model cannot be
specified beforehand, and only after a recourse to the methods of
kinetic theory can one estimate the friction coefficient to be imposed
in order to obtain a given viscosity \cite{Espanol1995pre, Marsh1997,
  Masters1999, Evans1999}.  As we will see, inclusion of more
sophisticated shear forces allows for a more direct connection with
Navier-Stokes hydrodynamics.

A third limitation of DPD as a mesoscale hydrodynamic solver is the
fact that the DPD model (in an identical manner as MD) is
\emph{hardwired to the scale}.  What we mean with this is that given a
physical problem, with a characteristic length scale, we may always
put a given number of dissipative particles and parametrize the model
in order to recover some macroscopic information (typically,
compressibilities and viscosity).  However, if one uses a different
number of particles for exactly the same physical situation, one
should start over and reparametrize the system again.  This is
certainly very different from what one would expect from a
Navier-Stokes solver, that specifies the equation of state and
viscosity irrespective of the scale, and one simply worries about
having a sufficiently large number of points to resolve the
characteristic length scales of the flow.  In other words, in DPD
there is no notion of \emph{resolution}, \emph{grid refinement}, and
\emph{convergence} as in CFD.  There have been attempts to restore a
\emph{scale free} property for DPD \cite{Espanol1998pre, Fuchslin2009,
  Arienti2011}, even for bonded interactions  \cite{Spaeth2010a}.  To
get this property, the parameters in the model need to depend on the
level of coarse-graining, but this is not specified in the original
model. Closely related to this lack of scaling is the fact that there
is no mechanism in the model to switch off thermal fluctuations
depending on the scale at which the model is operating.  On general
statistical mechanics grounds, thermal fluctuations should scale as
$1/\sqrt{N}$ where $N$ is the number of degrees of freedom
coarse-grained into one coarse-grained (CG) particle.  As the
dissipative particles represent larger and larger volume elements,
they should display smaller and smaller fluctuations.  But there is no
explicit volume or size associated to a dissipative particle.  This
problem is crucial, for instance, in the case of suspended colloidal
particles or in microfluidics applications where flow conditions and
the physical dimensions of the suspended objects or physical
dimensions of the operating device determine whether and, more
importantly, \emph{to what extent} thermal fluctuations come into
play.

Finally, another limitation of the DPD model is that it cannot sustain
temperature  gradients. Energy  in the  system is  dissipated and  not
conserved,  and  the   Brownian  dashpot  forces  of  DPD   act  as  a
thermostat.

\section{MDPD, EDPD, FPM}\label{Sec:MDPD}
During the years, the DPD  model has been \emph{enriched} in several
directions in order  to deal with all the above  limitations.  In this
Section we briefly review these enriched DPD models.

The many-body (or multi-body) dissipative particle dynamics (MDPD)
method stands for a modification of the original DPD model in which
the purely repulsive conservative forces of the classic DPD model are
replaced by forces deriving from a many-body potential; thus the
scheme is still covered by \Eqsref{dpd}--\eqref{gibbs}, but a
many-body $V(\{\rvec\})$ is substituted for \Eqref{V}.  The MDPD
method was originally introduced by Pagonabarraga and Frenkel
\cite{Pagonabarraga2001}, Warren \cite{Warren2001}, and independently
by Groot, \cite{footnote3} and subsequently modified and improved by
Trofimov \etal\ \cite{Trofimov2002}, reaching a level of maturity
\cite{Warren2003, Merabia2007, Ghoufi2010, Arienti2011, Ghoufi2013,
  Warren2013, AM16}.  The key innovation of the MDPD is the
introduction of a density variable $d_i=\sum_{j\neq i}W(r_{ij})$, as
well as a free energy $\psi(d_i)$ associated to each dissipative
particle.  Here $W(r)$ is a normalized bell-shaped weight function
that ensures that the density $d_i$ is high if many particles are
accumulated near the $i$-th particle.  The potential of interaction of
these particles is assumed to be of the form $V=\sum_i\psi(d_i)$
\cite{Warren2003b}.  This is a many-body potential of a form similar
to the embedded atom potential in MD simulations \cite{Daw1984,
  Daw1993}.  For multi-component mixtures the many-body potential may
be generalized to depend on partial local densities.

Despite its many-body character, the resulting forces are
  still pair-wise, and  implementation is straightforward.
  However, not all pair-wise force laws correspond to a many-body
  potential.  Indeed the existence of such severely constrains the
  nature of the force laws, and some errors have propagated into the
  literature (see discussion in \Refcite{Warren2013}).  In Appendix
  \ref{app:mdpd} we explore how the force law is constrained by the
  weight function $W(r)$.  The message is: if in doubt, always work from
  $V(\{\rvec\})$.

MDPD  escapes  the  straitjacket  of  mean-field  fluid  behaviour  by
modulating the thermodynamic  behaviour of the system  directly at the
interaction level between the particles.  This allows for more general
equations of state than in the  original DPD model, which is a special
case where the one-body terms  are linear in local densities.  Indeed,
one can easily engineer a van der  Waals loop in the equation of state
to accommodates vapor-liquid  coexistence.  But this in  itself is not
enough  to  stabilize  a  vapour-liquid interface,  since  one  should
additionally ensure  that the  cohesive contribution  is longer-ranged
than the repulsive contribution.  This  can be achieved for example by
using   different   ranges   for    the   attractive   and   repulsive
forces \cite{Warren2001, Warren2003}, or  modelling the  square gradient
term in the free energy \cite{Tiwari2006}.

The energy-conserving dissipative particle dynamics model (EDPD) was
introduced simultaneously and independently by Bonet Aval\'{o}s and
Espa\~nol \cite{Avalos1997, Espanol1997DPDE} as a way to extend the
DPD model to non-isothermal situation.  In this case, the key
ingredient is an additional internal energy variable associated to the
particles.  The behaviour of the model was subsequently studied
\cite{Avalos1999, Ripoll1998, Ripoll2000, Ripoll2005}.  The method has
been compared with standard flow simulations \cite{Abu-Nada2011,
  Yamada2011}, and recently a number of interesting applications have
emerged \cite{Chaudhri2009, Lisal2011}, including heat transfer in
nanocomposites \cite{Qiao2007}, shock detonations \cite{Maillet2011},
phase change materials for energy storage \cite{Rao2012}, shock
loading of a phospholipid bilayer \cite{Ganzenmuller2012}, chemically
reacting exothermic fluids \cite{Brennan2014, Moore2016},
thermoresponsive polymers \cite{Li2015}, and water solidification
\cite{Johansson2016}.

The fluid particle model (FPM) was devised as a way to overcome the
limitation of the simplistic friction forces in DPD
\cite{Espanol1997epl, Espanol1998pre}. The method introduced, in
addition to radial friction forces, shear forces that depend not only
on the approaching velocity but also on the velocity differences
directly. Shear forces have been reconsidered recently
\cite{Junghans2008}. The resulting forces are non-central and do not
conserve angular momentum.  In order to restore angular momentum
conservation a spin variable is introduced. Heuristically, the spin
variable is understood as the angular momentum relative to the center
of mass of the fluid particle.  The model has been used successfully
by Pryamitsyn and Ganesan \cite{Pryamitsyn2005} in the simulation of
colloidal suspensions, where each colloid is represented by just one
larger dissipative particle, an approach also used by Pan
\etal\ \cite{Pan2008}.

\section{DPD from top-down: The SDPD model}\label{Sec:topdown}
While MDPD is still isothermal and EDPD still uses conservative forces
too limited to reproduce arbitrary thermodynamics, the two enrichments
of a density variable and an internal energy variable introduced by
these models suggest a view of the dissipative particles as truly
thermodynamic subsystems of the whole system, consistently with the
local equilibrium assumption in continuum hydrodynamics.  There have
been a number of works trying to formalize this view of ``moving fluid
particles'' in terms of Voronoi cells of points moving with the flow
field \cite{Espanol1997}.  Flekk{\o}y \etal\ formulated a (semi)
bottom-up approach for constructing a model of fluid particles with
the Voronoi tessellation \cite{Flekkoy1999, Flekkoy2000}.  A
thermodynamically consistent Lagrangian finite volume discretization
of LLNS using the Voronoi tessellation was presented by Serrano and
Espa\~nol \cite{Serrano2001} and compared favourably
\cite{Serrano2002} with the models in \Refscite{Flekkoy1999} and
\cite{Flekkoy2000}.  While this top-down modeling based on the
Voronoi tessellation is grounded in a solid theoretical framework, it
has not found much application due, perhaps, to the computational
complexity of a Lagrangian update of the Voronoi tessellation
\cite{footnote1}.

In an attempt to simplify the Lagrangian finite Voronoi volume
discretization model, the smoothed dissipative particle dynamics
(SDPD) model was introduced shortly after \cite{Espanol2003SDPD},
based on its precursor \cite{Espanol1999PRL}.  SDPD is a
thermodynamically consistent particle model based on a particular
version of smoothed particle hydrodynamics (SPH) that includes thermal
fluctuations.  SPH is a mesh-free Lagrangian discretization of the
Navier-Stokes equations (NSE) differing from finite volumes, elements,
or differences in that a simple smooth kernel is used for the
discretization of space derivatives.  This leads to a model of moving
interacting point particles whose simulation is very similar to MD.
SPH was introduced in an astrophysical context for the simulation of
cosmic matter at large scales \cite{Lucy1977, Gingold78}, but has been
applied since then to viscous and thermal flows \cite{LiuLiu2003,
  Liu2010}, including multi-phasic flow \cite{Zhi-bin2016}.  An
excellent recent critical review on SPH is given by Violeau and Rogers
\cite{Violeau2016}.

In  the particular  SPH discretization  given by  SDPD of  the viscous
terms in the NSE, the resulting  forces have the same structure of the
shear friction forces  in the FPM.  By casting  the model within
the universal thermodynamically  consistent \GENERIC\ framework
 \cite{Ottinger2005}, thermal fluctuations  are introduced consistently
in SDPD by respecting an  exact fluctuation-dissipation theorem at the
discrete level.  Therefore,  SDPD (as opposed to SPH)  can address the
mesoscopic realm where thermal fluctuations are important.

The SDPD model consists on $N$ point particles characterized by their
positions and velocities ${\bf r}_i,{\bf v}_i$ and, in addition, a
thermal variable like the entropy $S_i$ (by a simple change of
variables, one can also use alternatively the internal energy
$\epsilon_i$ or the temperature $T_i$).  Each particle is understood
as a thermodynamic system with a volume ${\cal V}_i$ given by the
inverse of the density $d_i=\sum_i^NW(r_{ij})$, a fixed constant mass
$m_i$, and an internal energy $\epsilon_i=E(S_i,m_i,{\cal V}_i)$ which
is a function of the entropy of the particle, its mass (\ie\ number of
moles), and volume.  The functional form of $E(S,M,{\cal V})$ is
assumed, through the local equilibrium assumption, to be the same
function that gives the global thermodynamic behaviour of the fluid
system (but see below).  The equations of motion of the independent
variables are \cite{Espanol2003SDPD}
\begin{align}
d{\bf r}_i =&{\bf v}_idt 
\nonumber\\
m  d{\bf v}_i=& \sum_{j}\left[\frac{P_i}{d_i^2}
+\frac{P_j}{d_j^2}\right] F_{ij}{\bf r}_{ij}dt 
\nonumber\\
&- \frac{5\eta}{3}
\sum_j \frac{F_{ij}}{d_id_j}
\left({\bf v}_{ij}+{\bf e}_{ij}{\bf e}_{ij} \!\cdot\! {\bf v}_{ij}
\right)dt 
+m  d\tilde{\bf v}_i 
\nonumber\\
T_id{S}_i =& 
\frac{5\eta}{6}
\sum_j \frac{F_{ij}}{d_id_j}
\left({\bf v}_{ij}^2+({\bf e}_{ij} \!\cdot\! {\bf v}_{ij})^2
\right)dt 
\nonumber\\
&- 2\kappa \sum_j \frac{F_{ij}}{d_id_j}T_{ij}dt 
+ T_i d\tilde{S}_i
\label{sdefin}
\end{align}
Here, $P_i,T_i$ are the pressure and temperature of the fluid particle
$i$,  which  are  functions   of  $d_i,S_i$  through  the  equilibrium
equations  of state,  derived from  $E(S,M,{\cal V})$  through partial
differentiation.  Because the  volume  of a  particle  depends on  the
positions of  the neighbours, the  internal energy function  plays the
role of the potential energy $V$  in the original DPD model.  In addition,
${\bf v}_{ij}  = {\bf v}_{i}-{\bf v}_{j}$,  and $T_{ij}=T_i-T_j$.  The
function $F(r)$ is  defined in terms of the weight  function $W(r)$ as
$\boldsymbol{\nabla} W(r)  = -  {\bf r} F(r)$.   Finally, $d\tilde{\bf
  v}_i,d\tilde{S}_i$  are linear  combinations  of independent  Wiener
processes whose  amplitude is dictated by  the exact fluctuation-dissipation
theorem \cite{footnote2}.

It is easily shown that the above model conserves mass, linear
momentum and energy, and that the total entropy is a non-decreasing
function of time thus respecting the second law of thermodynamics.
The equilibrium distribution function is given by Einstein expression
in the presence of dynamic invariants \cite{Espanol1992}.  As the
number of particles increases, the resulting flow converges towards
the solution of the Navier-Stokes equations, by construction.

SDPD can be considered as the general version of the three models
MDPD, EDPD, FPM, discussed in \Secref{Sec:MDPD}, incorporating all
their benefits and none of its limitations.  For example, the pressure
and any other thermodynamic information is introduced as an input, as
in the MDPD model.  The conservative forces of the original model
become physically sounded pressure forces.  Energy is conserved and we
can study transport of energy in the system as in EDPD.  The transport
coefficients are input of the model (though, see below).  The range
functions of DPD have now very specific forms, and one can use the
large body of knowledge generated in the SPH community to improve on
the more adequate shape for the weight function $W(r)$ \cite{Liu2010}.
The particles have a physical size given by its physical volume and it
is possible to specify the physical scale being simulated.  One should
understand the density number of particles as a way of controlling the
\emph{resolution} of the simulation, offering a systematic `grid'
refinement strategy.  In the SDPD model, the amplitude of thermal
fluctuations scales with the size of the fluid particles: \emph{large}
fluid particles display smaller thermal fluctuations, in accordance
with the usual notions of equilibrium statistical mechanics.  While
the fluctuations scale with the size of the fluid particles, the
resultant stochastic forces on \emph{suspended} bodies are independent
of the size of the fluid particles and only depend on the overall size
of the object \cite{Vazquez-Quesada2009jcp}, as it should.

The SDPD model does not conserves angular momentum because the
friction forces are non-central.  This may be remedied by including an
extra spin variable as in the FPM as has been done by M\"uller
\etal\ \cite{Muller2015}. This spin variable is expected to relax
rapidly, more and more so as the size of the fluid particles
decreases.  For high enough resolution the spin variable is slaved by
vorticity. The authors of \Refcite{Muller2015} have shown, though,
that the inclusion of the spin variable may be crucial in some
problems where ensuring angular momentum conservation is important
\cite{Greenspan1968}.

In summary, SDPD can be understood as MDPD for non-isothermal
situations, including more realistic friction forces.  The SDPD model
has a similar simplicity as the original DPD model and its enriched
versions MDPD, EDPD, FPM.  It has been remarked \cite{Lei2015} that
SDPD does not suffer from some of the issues encountered in Eulerian
methods for the solution of the LLNS equations.  The SDPD model is
applicable for the simulation of complex fluid simulations for which a
\emph{Newtonian solvent} exists.  The number of studies using SDPD is
now growing steadily and range from microfluidics,\cite{Fan2006} and
nanofluidics \cite{Lei2015}, colloidal suspensions \cite{Bian2012,
  Vazquez-Quesada2015}, blood \cite{Moreno2013, Muller2014}, tethered
DNA \cite{Litvinov2011}, and dilute polymeric solutions
\cite{Litvinov2008, Litvinov2010, Litvinov2016}.  Also, it has also
been used for the simulation of fluid mixtures \cite{Thieulot2005,
  Thieulot2005a, Thieulot2005b, Petsev2016}, and viscoelastic flows
\cite{Vazquez-Quesada2009pre}.

Once SDPD is understood as a particle method for the numerical
solution of the LLNS equations of fluctuating hydrodynamics, the issue
of boundary conditions emerge. While there is an extensive literature
in the formulation of boundary conditions in the deterministic SPH
\cite{LiuLiu2003}, and in DPD \cite{Revenga1998, Revenga1999,
  Pivkin2005, Haber2006, Pivkin2006, Altenhoff2007, Henrich2007,
  Xu2009, Lei2011, Groot2012, Mehboudi2014}, the consideration of
boundary conditions in SDPD has been addressed only recently
\cite{Kulkarni2013, Gatsonis2014, Petsev2016}.

In SDPD, what you put is \emph{almost} what you get. The input
information is the internal energy of the fluid particles as a
function of density and entropy (or temperature), and the
viscosity. However, only in the high resolution limit, for a large
number of particles it is ensured the convergence towards the
continuum equations.  Therefore, for a finite number of particles
there will be always differences between the input viscosity and the
actual viscosity of the fluid and, possibly, between the input
thermodynamic behaviour of the fluid particle and the bulk system.
These differences could be attributed to numerical ``artifacts'' of
the particle model, similar to discretisation errors that arise in
CFD. Often the worst effects of these artifacts can be eliminated by
using renormalized transport coefficients from calibration
simulations.  This is similar, for instance, to the way that
discretisation errors in lattice Boltzmann are commandeered to
represent physics, improving the numerical accuracy of the scheme
\cite{Anc94}.  In this context the availability of a systematic grid
refinement strategy for SDPD is clearly highly beneficial.

\subsection{Internal variables}\label{internal}
The SDPD model is obtained from the discretization of the continuum
Navier-Stokes equations.  Of course, any other continuum equations
traditionally used for the description of complex fluids can also be
discretized with the same methodology.  In general, these continuum
models for complex fluids typically involve \emph{additional
  structural or internal variables}, usually representing
mesostructures, that are coupled with the conventional hydrodynamic
variables \cite{Kroger2004, Ottinger2005}.  The coupling of
hydrodynamics with these additional variables renders the behaviour of
the fluid non-Newtonian and complex.  For example, polymer melts are
characterized by additional conformation tensors, colloidal
suspensions can be described by further concentration fields, mixtures
are characterized by several density fields (one for each chemical
species), emulsions are described with the amount and orientation of
interface, \etc.

All these continuum models rely on the hypothesis of local equilibrium
and, therefore, the fluid particles are regarded as thermodynamic
subsystems. Once the continuum equations are discretized in terms of
fluid particles (Lagrangian nodes) with associated additional
structural or order parameter variables, the resulting fluid particles
are ``large'' portions of the fluid.  The scale of these fluid
particles is \emph{supra-molecular}.  This allows one to study larger
length and time scales than the less coarse-grained models where the
mesostructures are represented explicitly through additional
interactions between particles (\ie\ chains for representing polymers,
spherical solid particles to represent colloid, different types of
particles to represent mixtures).  The price, of course, is the need
for a deep understanding of the physics at this more coarse-grained
level, which should be adequately captured by the continuum equations.

For example, in order to describe polymer solutions, we may take a
level of coarse graining in which every fluid particle contains
already many polymer molecules.  This is a more coarse-grained model
than describing viscoelasticity by joining dissipative particles with
springs \cite{Somfai2006}. The state of the polymer molecules within a
fluid particle may be described either with the average end-to-end
vector of the molecules \cite{tenBosch1999, Ellero2003}, or
with a conformation tensor \cite{Vazquez-Quesada2009pre}. In this
latter case, the continuum limit of the model leads to the Olroyd-B
model of polymer rheology.  Another example where the strategy of
internal variables is successful is in the simulation of
mixtures. Instead of modeling a mixture with two types of dissipative
particles as it is usually done in DPD, one may take a
thermodynamically consistent view in which each fluid particle
contains the concentration of one of the species, for example
\cite{Thieulot2005, Thieulot2005a, Li2015a, Petsev2016}.  Chemical
reactions can be implemented by including as an internal degree of
freedom an extent of reaction variable \cite{Brennan2014}.

\section{DPD from bottom-up}\label{Sec:bottomup}
The SDPD model \cite{Espanol2003SDPD}, or the Voronoi fluid particle
model \cite{Serrano2001}, are top-down models which are, essentially,
Lagrangian discretizations of fluctuating hydrodynamics.  These models
are the bona fide connection of the original DPD model with continuum
hydrodynamics.  However, the connection of the model with the
microscopic level of description is less clear. Ideally, one would
like to fulfill the program of coarse-graining, in which starting from
Hamilton's equations for the atoms in the system, one derives closed
equations for a set of CG variables that represent the system in a
fuzzy impressionistic way.

Coarse graining of  a molecular system requires a  clear definition of
the mapping  between the microscopic  and CG degrees of  freedom. This
mapping is usually well defined when the atoms are bonded, as happens
inside complex molecules like proteins and other polymer molecules, or
in solid  systems. In this  case, one can  choose groups of  atoms and
look  at,  for  example, the  center  of  mass  of  each group  as  CG
variables. For unbonded  atoms as those occurring in  a fluid system,
the main  problem is that grouping  atoms in a system  where the atoms
may  diffuse away  from  each  other is  a  tricky  issue. We  discuss
separately the strategies  that have been followed in  order to tackle
the coarse-graining of both, unbonded and bonded atoms.

\subsection{DPD for unbonded atoms}
The derivation of the equations of hydrodynamics from the underlying
Hamiltonian dynamics of the atoms is a well studied problem that dates
back to Boltzmann and the origins of kinetic theory \cite{Irving1950,
  Grabert1982}.  It is a problem that still deserves attention for
\emph{discrete} versions of hydrodynamics \cite{DelaTorre2011,
  Espanol2009, Espanol2009c, EspanolDonev2015}, which is what we need
in order to simulate hydrodynamics in a computer.  These latter works
show how an \emph{Eulerian} description of hydrodynamics can be
derived from the Hamiltonian dynamics of the underlying atoms, by
defining mass, momentum, and energy of cells which surround certain
points fixed in space.  However, \emph{Lagrangian} descriptions in
which the cells ``move following the flow'', are much more tricky to
deal with. Typically, two types of groupings of fluid molecules have
been considered, based on the Voronoi tessellation or on spherical
blobs.

An early attempt to construct a Voronoi fluid particle out from the
microscopic level was made by Espa\~nol \etal\ \cite{Espanol1997}. The
Voronoi centers were moved according to the forces felt by the
molecules inside the cell in the underlying MD simulation.  An
effective excluded volume potential was obtained from the radial
distribution function of the Voronoi centers.  The method was
revisited by Eriksson \etal\ \cite{Eriksson2009b} who observed
``molecular unspecificity'' of the Voronoi projection, in the sense
that very different microscopic models give rise to essentially the
same dynamics of the cells.  In earlier work \cite{Eriksson2008}, a
force covariance method, essentially the Einstein-Helfand route to
compute the Green-Kubo coefficients \cite{Kauzlaric2011}, was
introduced in order to compute the friction forces under the DPD
ansatz.  The results are disappointing as these authors showed that
the dynamics of the CG particles with the forces of the DPD model
measured from MD for a Lennard-Jones system were not consistent with
the MD results themselves.

More recently, Hadley and McCabe \cite{Hadley2010} propose to group
water molecules into beads through the $K$-means algorithm
\cite{Macqueen1967}.  The algorithm considers a number of beads with
initially given positions and construct their Voronoi tessellation.
The water molecules inside each Voronoi cell have a center of mass
that does not coincide with the bead position.  The bead position is
then translated on top of the center of mass and a retessellation is
made again, with a possibly different set of water molecules
constituting the new bead.  The procedure is repeated until
convergence.  At the end, one has centroidal Voronoi cells in which
the bead position and the center of mass of the water molecules inside
the Voronoi cell coincide.  The $K$-means algorithm gives for every
microstate (coordinates of water molecules) the value of the
macrostate (coordinates of the beads) and, therefore, provides a
rule-based CG mapping.  Unfortunately, there is no analytic function
that captures this mapping and, therefore, it is not possible to use
the theory of coarse-graining to rigorously derive the evolution of
the beads.  The strategy by Hadley and McCabe is to construct the
radial distribution function and infer from it the pair potential.
Recently, Izvekov and Rice \cite{Izvekov2015} have also considered
this procedure in order to compute both, the conservative force and
the friction force between beads by extracting this information from
force and velocity correlations between Voronoi cells. They find that
very few molecules per cell are sufficient to obtain Markovian
behaviour.

Instead of using Voronoi based fluid particles, Voth and co-workers
consider a sphere (termed a \BLOB) and move the sphere according to
the forces experienced by the center of mass of the molecules inside
it \cite{Ayton2004a}. The dynamics of the \BLOB\ is then modeled in
order to reproduce the time correlations of the \BLOB. Subsequently a
system of $N$ Brownian \BLOB s is constructed in order to reproduce
the above correlations.

Recently, another attempt to obtain DPD from the underlying MD has
been undertaken by Lei \etal\ \cite{Lei2010} by using the rigorous
approach of the theory of coarse-graining.  However, in order to
construct the ``fluid particles'' these authors constraint a
collection of Lennard-Jones atoms to move bonded, by maintaining a
specified radius of gyration. The fluid no longer is a simple atomic
fluid but rather a fluid made of complex ``molecules'' (the atomic
clusters constrained to have a radius of gyration) whose rheology is
necessarily complex.

Our impression is that we still have not solved satisfactorily the
problem of deriving from the microscopic dynamics the dynamics of CG
particles that capture the behaviour of a simple fluid made of
\emph{unbonded} atoms.  Work remains to be done in order to define the
proper CG mapping for a fully satisfactory bottom-up model for
Lagrangian fluid particles representing a set of few unbonded atoms or
molecules ``moving coherently''.

\subsection{ DPD for bonded atoms}
\begin{figure}[t]
    \centering
    \includegraphics[scale=0.2,angle=90]{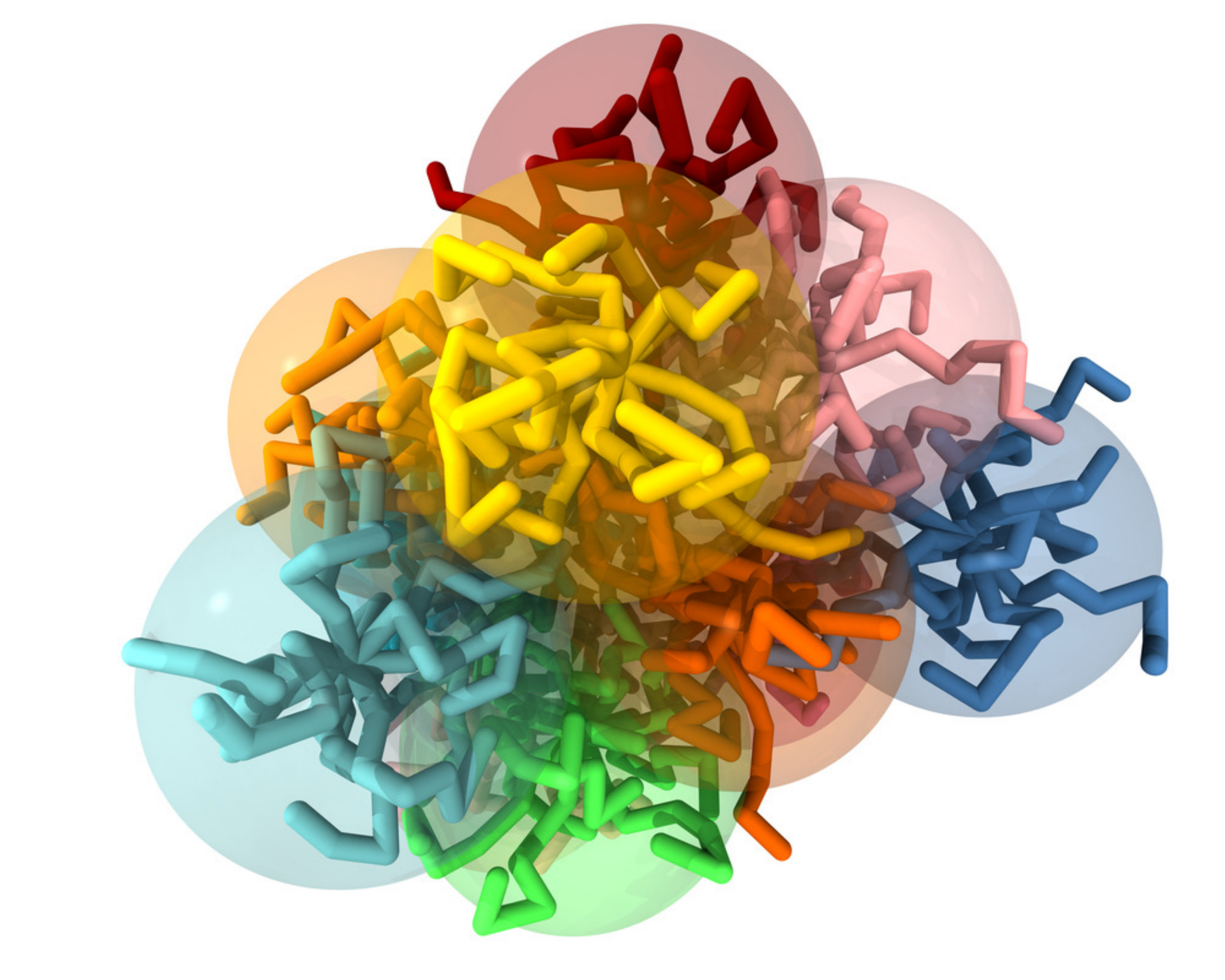}
    \caption{Star polymer molecules (in different colors) in a melt
      are coarse-grained at the level of their centers of mass. The
      resulting model is a blob model of the DPD type
      \cite{Hijon2010}.}
    \label{Fig.star}
\end{figure}

When the atoms are bonded and belong to definite groups where the
atoms do not diffuse away from each other, the CG mapping is well
defined, usually through the center of mass variables.  In
\Figref{Fig.star} we show a star polymer melt in which each molecule
is coarse-grained by its center of mass, leading to a blob or bead
description \cite{Hijon2010}. The important question is how are the CG
interactions between the blobs. Two CG approaches, static and dynamic,
have been pursued, depending on the questions one wishes to answer.

\emph{Static CG} is concerned with approximations to the exact
potential of mean force that gives, formally, the equilibrium
distribution function of all the CG degrees of freedom.  Radial
distributions, equations of state, \etc.  are the concern of static
coarse graining.  There is a vast literature in the construction of
the potential of mean force for CG representations of complex fluids
\cite{Reith2001, Likos2001, Voth2009}, and complex molecules
\cite{Milano2005a, Noid2013, Lopez2014}. Despite these efforts, there
is still much room for improvement in the thermodynamic consistency
for the modeling of the potentials of mean force
\cite{Reinier2001}. If one uses the CG potential for the motion of the
CG degrees of freedom, the resulting dynamics is unrealistically fast,
although this may be in some cases convenient computationally.

\emph{Dynamic CG}, on the other hand, focuses on obtaining, in
addition to CG potentials, approximations to the friction forces
between CG degrees of freedom.  Within the theoretical framework of
Mori-Zwanzig approach, it is possible to obtain in general the
dynamics of the CG degrees of freedom from the underlying Hamiltonian
dynamics.  The first attempt to derive the DPD model from the
underlying microscopic dynamics was given by Espa\~nol for the simple
case of a one dimensional harmonic lattice
\cite{Espanol1996pre-harmonic}.  The center of mass of groups of atoms
were taken as the CG variables and Mori's projection method was used.
Because this system is analytically soluble, a flaw in the original
derivation could be detected, and an interesting discussion emerged on
the issue of non-Markovian effects in solid systems \cite{Cubero2005,
  Cubero2005a, Hijon2006, Cubero2008a, Hijon2008}.

By following Schweizer \cite{Schweizer1989}, Kinjo and Hyodo
\cite{Kinjo2007} obtained a formal equation for the centers of mass of
groups of atoms.  The momentum equation contains three forces, a
conservative force deriving from the exact potential of mean force, a
friction force and a random force.  By \emph{modeling} the random
forces the authors of \Refcite{Kinjo2007} showed that this equation
encompass both, the BD and DPD equations.  However, to consider the
procedure in \Refcite{Kinjo2007} a \emph{derivation} of DPD, it is
necessary to specify the conditions under which one obtains BD instead
of DPD (or \viceversa).  This was not stated by Kinjo and Hyodo.  The
crucial insight is that BD appears when ``solvent'' is eliminated from
the description, this is, some (the majority) of the atoms are not
grouped and are instead described as a passive thermal bath (or
implicit solvent).  The friction force in this case is proportional to
the velocity of the particles, and the momentum of the CG blobs is not
conserved.  On the other hand, a DPD description appears when
\emph{all} the atoms are partitioned into disjoint groups. In this
case, the conservation of momentum induced by Newton's third law at
the microscopic level leads to a structure of the friction forces
depending on \emph{relative} velocities of the particles.  A
derivation of the equations of DPD from first principles taking into
account linear momentum conservation was presented by Hijon
\etal\ \cite{Hijon2010}.  The position-dependent friction coefficient
was given in terms of a Green-Kubo expression that could be evaluated,
under certain simplifying assumptions, directly from MD simulations,
within the same spirit of an early derivation of Brownian Dynamics for
a dimer representation (non-momentum conserving) of a polymer by
Akkermans and Briels \cite{Akkermans2000}.  The general approach was
preliminarly tested for a system of star polymers (as those in
\Figref{Fig.star}).  A subsequent thorough study of this star polymer
problem by Karniadakis and co-workers \cite{Li2014} has shown that the
introduction of an intrinsic spin variable for each polymer molecule
seems to be necessary at low concentrations in order to have an
accurate representation of the MD results.  The approach in
\Refcite{Hijon2010} has been labeled by Li \etal\ \cite{Li2014} as the
MZ-DPD approach, standing for Mori-Zwanzig dissipative particle
dynamics.  Other complex molecules (neopentane, tetrachloromethane,
cyclohexane, and n-hexane) have been also considered
\cite{Deichmann2014} within the MZ-DPD approach with interesting
discussion on the validity of non-Markovian behaviour (more on this
later).  A slightly more general approach for the derivation of MZ-DPD
equations has been given by Izvekov \cite{Izvekov2015}.  Very
recently, Espa\~nol \etal\ \cite{Espanol2016} have formulated from
first principles the dynamic equations for an \emph{energy conserving}
CG representation of complex molecules.  This work gives the
microscopic foundation of the EDPD model for complex molecules
(involving bonded atoms only).

\subsection{Non-Markov effects}
The rigorous coarse-graining in which centers of mass of groups of
atoms are used as CG variables relies on a basic and fundamental
hypothesis, which is the separation of time scales of the evolution of
the CG variables and ``the rest'' of variables in the system. More
accurately, the separation of time scales refers to the existence,
\emph{in the evolution of the CG variables themselves} of two
well-defined scales, a large amplitude slow component, and a small
high frequency component that can be modeled in terms of white noise.
The dynamics of the CG variables can then be approximately described
by a non-linear diffusion equation in the space spanned by the CG
variables \cite{Green1952, Zwanzig1961}. This separation of time
scales does not always exist, either because the groups of atoms are
small and the centers of mass momenta evolve in the same time scales
as the forces (due to collisions with atoms of other groups)
\cite{Deichmann2014}, or because of the existence of coupled slow
processes not captured by the selected CG variables.  When this
happens, one strategy is to tweak the friction and simply fit
frictions to recover the time scales.  Gao and Fang used this approach
in order to coarse grain a water molecule to one site-CG particle
\cite{Gao2011}.  Another strategy is to enlarge the set of CG
variables with the hope that the new set will be Markovian.  Briels
\cite{Briels2009} addresses specifically the problem of CG in polymers
and introduces transient forces to recover a Markovian description.
Davtyan, Voth, and Anderson \cite{Davtyan2015} have considered the
introduction of ``fictitious particles'' in order to recover the CG
dynamics observed from MD. The fictitious particles are just a simple
and elegant way to model the memory kernel in a particularly intuitive
way.  If the strategy to increase the dimension of the CG state space
does not work yet, it is still possible to formulate from microscopic
principles formal non-Markovian models and to extract information
about the memory kernel from MD \cite{Yoshimoto2013}.  However, in the
absence of separation of time scales, the computational effort
required to get from MD the memory kernel makes the whole strategy of
bottom-up coarse graining inefficient.  Note that the advantage of a
bottom-up strategy for coarse graining is that one needs to run
\emph{relatively short} MD simulations to get the information
(Green-Kubo coefficients) that is used in the dynamic equations
governing much larger time scales.  If one needs to run long MD
simulation of the microscopic system to get the CG information, we
have already solved the problem by brute force in the first place!

\subsection{Electrostatic interactions}
In many situations, one is interested in the consequential effects of
charge separation.  This is particularly so for aqueous systems where
the relatively high dielectric permittivity of water means that ion
dissociation readily occurs.  The relevance lies not only in the
structural and thermodynamic properties of ionic surfactants and
polyelectrolytes \etc, \cite{Sindelka2014, LLP16} but is also
motivated by a burgeoning interest of electrokinetic
phenomena \cite{Pagonabarraga2010, Smiatek2012, Maduar2015,
  Sepehr2016}.

An important point to make is that some relevant electrostatic effects
simply cannot be captured in a short-range interaction (DPD-like or
otherwise).  For example the bare electrostatic energy of a charged
spherical micelle of aggregation number $N$ scales as $N^2/R \sim
N^{5/3}$, where $R\sim N^{1/3}$ is the micelle radius.  This
electrostatic energy cannot be captured in either a volume ($\sim N$)
or surface ($\sim N^{2/3}$) term.  To be faithful to this physics
therefore, one has in some way to incorporate long-range Coulomb
interactions explicitly into the DPD model.  This area was pioneered
by Groot, who used a field-based method \cite{Groot2003}.  Since then,
more standard Ewald methods have also been used \cite{GMV+06, WVA+13},
and in principle any fast electrostatics solver developed for MD could
be taken over into the DPD domain.  One important caveat is that with
soft particles (ie no hard core repulsion) the singularity of Coulomb
interactions needs to be tamed through the use of smoothed charge
models \cite{Groot2003, GMV+06, WVA+13, Warren2014}.  Adding
electrostatics is certainly computationally expensive, often halving
the speed of DPD codes. This drastic slow-down offsets the advantage
of the DPD methods when compared to more traditional coarse-graining
methods.

Related to the problem of charge separation is the question of
dielectric inhomogeneities (image charges \etc), such as encountered
in oil/water mixtures and at interfaces.  In field-based methods this
can be resolved by having a local-density-dependent dielectric
permittivity \cite{Groot2003}.  Alternatively one can introduce an
explicitly polarisable solvent model \cite{PP14, Peter2015a}.  Note that
an electric field in the presence of a dielectric inhomogeneity
induces reaction forces on the \emph{uncharged}
particles \cite{Groot2003}.  In a field-based method, this is quite
complicated to incorporate in a rigorous way.  For an explicitly
polarisable solvent model, these induced reaction forces of course
arise naturally and are automatically captured.

\section{Systems studied with DPD}\label{Sec:applications}
The number of  systems and problems that have been  addressed with DPD
or  its variants  is enormous  and  we do  not pretend  to review  the
extensive literature  on the  subject.  Nevertheless, to  illustrate the
range  and  variety  of  different  applications  of  DPD  we  give  a
necessarily brief survey of the field. A general trend observed in the
application side, is the shift from the original DPD model, of ``balls
and springs'' models,  towards more specific atomistic  detail, in the
line  of  MZ-DPD,  or  semi-bottom-up  DPD  (with structure  based  CG
potentials and fitted friction).

\emph{Colloids:} A recent review on the simulation of colloidal
suspensions with particle methods, including DPD, can be found in
\cite{Bolintineanu2014}.  The first application of DPD to a complex
fluid was the simulation of colloidal rheology by Koelman and
Hooggerbruge \cite{Koelman1993}.  Since then, a large number of works
have addressed the simulation of colloidal suspensions, with a variety
of approaches to represent the solute.  Typically, a colloidal
particle is constructed out of dissipative particles that are moved
rigidly \cite{Koelman1993, Li2008}, or connected with springs
\cite{Laradji2004, Phan-Thien2014}. Arbitrary shapes may be considered
in this way \cite{Boek1997}, as well as confinement due to walls
\cite{Gibson1999, Li2008}.  As a way to bypass the need to update the
relatively large number of solid particles, some approaches represent
each colloidal particle with a single dissipative particle
\cite{Pryamitsyn2005, Pan2008, Pan2010}, leading to minimal spherical
blob models for the colloids.  These simplified models for the solute
require the introduction of shear forces of the FPM type.
Representing a colloidal particle with a point particle is a strategy
also used in minimal blob models in Eulerian CFD methods for
fluctuating hydrodynamics \cite{BalboaUsabiaga2014}.  A core can be
added in order to represent hard spheres with finite radii,
supplemented with a dissipative surface to mimic boundary conditions
\cite{Whittle2010}, and still retain the one-particle-per-colloid
scheme.  Although general features show semi-quantitative agreement
with experimental results \cite{Whittle2010}, other simulation
techniques like Stokesian Dynamics, and theoretical work, it is clear
that getting more detailed physics of colloid-colloid interactions and
colloid-solvent interaction (either through a MZ-DPD approach or by
phenomenologically including boundary layers and top-down
parametrization) may be beneficial to the field.

\emph{Blood:} A colloidal system of obvious biological interest is
blood. Blood has been simulated with DPD \cite{Fedosov2011}, and more
recently with SDPD \cite{Moreno2013, Muller2014, Katanov2015}.  Two
recent reviews \cite{Li2013, Ye2015} discuss the modeling of blood
with particle methods.  Multi-scale modeling (\ie\ MZ-DPD) seems to be
crucial to capture platelet activation and thrombogenesis
\cite{Zhang2014}.

\emph{Polymers: } An excellent recent review on coarse-graining of
polymers is given by Padding and Briels \cite{Padding2011}.  Below the
entanglement threshold Rouse dynamics holds and this is well satisfied
in a DPD polymer melt \cite{Spenley2000}.  Above the threshold,
entanglements are a necessary ingredient in polymer melts.  Because
the structure based CG potential between the blobs are very soft, it
is necessary to include a mechanism for entanglement explicitly. This
is one example in which the usual simple schemes to treat the
many-body potential (through pair-wise interactions) fails
dramatically.  There are several methods to include entanglements:
Padding and Briels \cite{Padding2001, Padding2002} introduced the
elastic band method for coarse-grained simulations of polyethylene.
Another alternative to represent entanglements is to use the Kumar and
Larson method \cite{Kumar2001, Goujon2008, Yamanoi2011} in which a
repulsive potential between bonds linking consecutive blobs is
introduced.  Finally, entanglements can be enforced in a simpler way
by hard excluded volume LJ interactions \cite{Symeonidis2005}, or
through suitable criterion on the stretching of two bonds and the
amount of impenetrability of them \cite{Nikunen2007}.

Beyond scaling properties, effort has been directed towards a
chemistry detailed MZ-DPD methodology, by using structure based CG
effective potentials and either fitting the friction coefficient
\cite{Guerrault2004, Lahmar2007, Maurel2012, Maurel2015}, or obtaining
the dissipative forces from Green-Kubo expressions \cite{Trement2014}.
In general, one can take advantage of systematic static
coarse-graining approaches, like those for heptane and toluene
\cite{Dunn2015}, to be directly incorporated to DPD.  Very recently,
new Bayesian methods for obtaining the CG potential \emph{and}
friction are being considered \cite{Dequidt2015, Solano2016} (on
pentane).  The ultimate goal of all these microscopically informed
approaches is to predict rheological properties as a function of
chemical nature of the polymer system with a small computational
cost. As mentioned earlier, whatever improvement in the construction
of CG potentials will be highly beneficial also for the construction
of dynamic CG models. In this respect, the work on \emph{analytical}
integral equation approach of Guenza and co-workers \cite{McCarty2014}
for obtaining the CG potential in polymer systems that ensures both,
structural properties \emph{and} thermodynamic behaviour seems to be
very promising.

We perceive a powerful trend towards more microscopically informed DPD
able to express faithfully the chemistry of the system.  This trend is
important when considering hierarchical multi-scale methods in which
MD information is transferred to a dynamic CG DPD model, the DPD model
is evolved in order to get topology and equilibrium states much faster
than MD, and then a back-mapping fine grained procedure recovers
microscopic states able to be evolved again with MD \cite{Chen2006,
  Santangelo2007, Gavrilov2015}.

Other complex fluid systems involving polymers have been considered.
An early work is the study of adsorption of colloidal particles onto a
polymer coated surface \cite{Gibson1999}.  Polymer brushes are
reviewed by Kreer \cite{Kreer2016}.  Self assembly of giant
amphiphiles made of a nanoparticle with tethered polymer tail has been
considered recently \cite{Ma2015}.  Polymer membranes for fuel cells
have been considered by Dorenbos \cite{Dorenbos2015}. Polymer
solutions simulated with DPD obey Zimm theory that includes
hydrodynamic interactions \cite{Jiang2007}.  Polymer solutions have
also been studied with SDPD observing Zimm dynamics
\cite{Litvinov2008}.

\emph{Phase separating fluids:} In polymer mixtures, the
$\chi$-parameter mapping introduced by Groot and Madden
\cite{Groot1998} has been phenomenally popular because it links to
long-established polymer physical chemistry (there are tables of
$\chi$-parameters for instance, and a large literature devoted to
calculating $\chi$-parameters \abinitio).  This has helped incorporate
chemical specificity in DPD from solubility parameters
\cite{Maiti2004b, Liyana-Arachchi2015}.  It is also known that
$\chi$-parameters can be composition dependent (PEO in water is the
notorious example).  This can be accommodated within the MDPD
approach. Akkermans \cite{Akkermans2008} presents a first principles
coarse-graining method that allows to calculate the excess free energy
of mixing and Flory-Huggins $\chi$-parameter.  A related effort is
given by Goel \etal\ \cite{Goel2014a}.

DPD has been very successful in identifying mechanisms in phase
separation: Linear diblock copolymer spontaneously form a
mesocopically ordered structure (lamellar, perforated lamellar,
hexagonal rods, micelles) \cite{Groot1998}. DPD is capable to predict
the dynamical pathway towards equilibrium structures and it is
observed that hydrodynamic interactions play an important role in the
evolution of the mesophases \cite{Groot1999}.  Domain growth and phase
separation of binary immiscible fluids of differing viscosity was
studied in \cite{Novik2000}.  New mechanisms via inertial hydrodynamic
bubble collapse for late-stage coarsening in off-critical vapor-liquid
phase separation have been identified \cite{Warren2001}.  The effect
of nanospheres in the mechanisms for domain growth in phase separating
binary mixture has been considered by Laradji and Hore
\cite{Laradji2004}.

\emph{Drop dynamics:} A particular case of phase separating fluids is
given by liquid-vapour coexistence giving rise to
droplets. Surface-confined drops in a simple shear was studied in an
early work \cite{Jones1999}.  Pendant drops have been studied with
MDPD \cite{Warren2003}, while oscillating drops \cite{Liu2006}, and
drops on superhydrophobic substrates \cite{Wang2015}, have also been
considered.

\emph{Amphiphilic systems:} An early review of computer modeling of
surfactant systems is by Shelley and Shelley \cite{Shelley2000}. A
more recent review on the modeling of pure membranes and lipid-water
membranes with DPD is given by Guigas \etal\ \cite{Guigas2011}.
Coarsening dynamics of smectic mesophase of amphiphilic species for a
minimal amphiphile model was studied by Jury \etal\ \cite{Jury1999}
and mesophase formation in pure surfactant and solvent by Prinsen
\etal\ \cite{Prinsen2002}.  More microscopic detail has been included
by Ayton and Voth \cite{Ayton2002} with DPD model for CG lipid
molecules that self assembly, a problem also considered by Kranenburg
and Venturoli \cite{Kranenburg2003}.  Effort towards more realistic
parametrization for lipid bilayers was given by Gao
\etal\ \cite{Gao2007}.  Prior to this Li \etal\ \cite{Li2004}
formulated a conservative force derived from a bond-angle dependent
potential that allowed to consider different types of micellar
structures.  Microfluidic synthesis of nanovesicles was considered by
Zhang \etal\ \cite{Zhang2015}.  Simulations of micelle-forming systems
have also been reported \cite{VLN13, JSJ+16}.

\emph{Oil industry:} DPD simulations have also addressed problems in
the oil industry, from oil-water-surfactant dynamics
\cite{Rekvig2004}, and water-benzene-caprolactam systems
\cite{Shi2015}, to aggregate behavior of asphaltenes in heavy crude
oil \cite{Zhang2010}, or the orientation of asphaltene molecules at
the oil-water interface \cite{Ruiz-Morales2015}.

\emph{Biological membranes:} A review of mesoscopic modeling of
biological membranes was given by Venturoli
\etal\ \cite{Venturoli2006}.  Groot and Rabone \cite{Groot2001}
presented one of the first applications of DPD to the modeling of
biological membranes and its disruption due to nonionic
surfactants. Sevink and Fraaije \cite{Sevink2014} devised a
coarse-graining of a membrane into a DPD model in which the solvent
was treated implicitly.  Amphiphilic polymer coated nanoparticles for
assisted drug delivery through cell membranes has been recently
studied \cite{Zhang2015a, Zhang2015b}.  The diffusion of membrane
proteins has been considered iby Guigas and Weiss \cite{Guigas2015}.
 
\emph{Biomolecular modeling:} The CG modeling of complex biomolecules
with a focus on static properties has been addressed in the excellent
review by Noid \cite{Noid2013}.  Pivkin \etal\ \cite{Peter2015a} have
modeled proteins with DPD force fields, which competes with the
Martini force field \cite{Marrink2013}.

\emph{Inorganic  materials:}  DPD has  also  been  used for  the  CG
modeling of solid  inorganic materials. Coarse-grained representation
of  graphene turns  out  to  be essential  for  study  of large  scale
resonator technology \cite{Kauzlaric2011, Kauzlaric2011jcp}.

\section{Conclusions}\label{Sec:conclusions}
The DPD model is a tool for simulating the mesoscale.  The model has
evolved since its initial formulation towards enriched models that,
while retaining the initial simplicity of the original, are now linked
strongly to either the microscopic scale or the macroscopic continuum
scale.  In many respects, the original DPD model of
\Figref{Fig.Dashpot} is a toy model and one can do much better by
using these refined models.  In this Perspective, we wish to convey
the message that DPD has a dual role in modeling the mesoscale. It has
been used as a way to simulate, on one hand, coarse-grained (CG)
versions of complex molecular \emph{objects} and, on the other hand,
\emph{fluctuating fluids}.  While the first type of application,
involving atoms bonded by their interactions, has a solid ground on
the theory of coarse graining, there is no such a \emph{microscopic}
basis for DPD as a fluid solver.  The best we can do today is to
descend from the continuum theory and to formulate DPD as a Lagrangian
discretization of fluctuating hydrodynamics, leading to the SDPD
model.

Therefore  as  DPD simulators  we  are  faced with  three  alternative
strategies:

\emph{\#1 Bottom-up MZ-DPD:} When dealing with molecular objects made
of bonded atoms, we may formulate an appropriate CG mapping and
construct the DPD equations of motion with momentum conserving forces
\cite{Hijon2010}.  These equations contain the potential of mean force
generating conservative forces and position-dependent friction
coefficient, with explicit microscopic formulae: the potential of mean
force is given by the configuration dependent free energy function,
and the position dependent friction coefficient tensor is given by
Green-Kubo expressions.  Both quantities are given in terms of
expectations \emph{conditional} on the CG variables and are,
therefore, many-body functions.  These are not, in
general, directly computable due to the curse of dimensionality.  One
needs to formulate simple and approximate models (usually pair-wise
with, perhaps, bond-angle and torsion effects) in order to represent
the complex functional dependence of these quantities.  Together with
the initial selection of the CG mapping, finding suitable functional
forms is the most delicate part of the problem.  Once this simple
functional models are selected, constrained MD simulations
\cite{Akkermans2000, Hijon2010}, or optimization methods
\cite{Noid2013, Brini2013, Lopez2014, Dequidt2015}, may be used to
obtain the CG potential.

The existence of a framework to derive dynamic CG models from
bottom-up is a highly rewarding intellectual experience with a high
practical value because 1) it provides the \emph{structure} of the
dynamic equations, and 2) signals at the crucial points where
approximations are required.  The MZ-DPD approach is, in our view, an
important breakthrough in the field, as it connects the well
established world of \emph{static coarse-graining} with the DPD world
\cite{Noid2013}. In this way, it provides a framework for accurately
addressing the CG \emph{dynamics}.  However, the usefulness to follow
the program by the book is not always obvious due to the large effort
in obtaining the objects form MD.  In this case, one would go to the
next strategy.

\emph{\#2 Parametrization of DPD:} We may insist on a particularly
simple form of linear repulsive forces and simple friction
coefficients (like the ones in the original/cartoon DPD model) and fit
the parameters to whatever property of the system one wants to
correctly describe (for example, the compressibility).  Nowadays, we
advise caution with this simple approach because, usually, many other
properties of the system go wrong.  The simple DPD linear forces are
not flexible enough in many situations.  However, from what we have
already learned from microscopically informed MZ-DPD in the previous
\#1 strategy, we may give ourselves more freedom in selecting the
functional forms (as in MDPD) for conservative and friction forces and
have more free parameters to play with.  Once it is realized that the
potential between beads or blobs in DPD is, in fact, the potential of
mean force, one can use semi-bottom-up approaches in which the
potential of mean force is obtained from first principles, while the
DPD friction forces are fitted to obtain the correct time scales
\cite{Lyubartsev2002b, Guerrault2004, Lahmar2007}.  Although this
strategy is less rigorous, it may be more practical in some cases.

The \#1 bottom-up MZ-DPD strategy above  has not been yet successful when
the   interactions  of   atoms  or   molecules  in   the  system   are
\emph{unbonded},  allowing two  molecules that  are initially  close
together  to diffuse  away from  each other.   These are  the kind  of
interactions present in a fluid  system.  The main difficulty seems to
be in  the \emph{Lagrangian} nature  of a fluid particle that makes
the CG mapping not obvious.  
Although some  attempts have  been taken  in order  to derive  DPD for
fluid systems with unbonded interactions,  we believe that the problem
is not  yet solved.   However, for  these systems  one may  regard the
dissipative  particles   as  truly   fluid  particles   (\ie\   small
thermodynamic systems  that move with the  flow).  We are lead  to the
third strategy.

\emph{\#3 Top-down DPD:} Assume that we know that a particular field
theory describes the complex fluid of interest at a macroscopic scale
(Navier-Stokes for a Newtonian fluid, for example).  Then one may
discretize the theory on moving Lagrangian points according to the SPH
mesh-free methodology.  The Lagrangian points may be interpreted as
fluid particles.  If we perform this discretization within a
thermodynamically consistent framework like
\GENERIC\ \cite{Ottinger2005}, thermal fluctuations are automatically
determined correctly \cite{Petsev2016}, allowing to address the
mesoscale.  This strategy leads to enriched DPD models (SDPD is an
example corresponding to Navier-Stokes hydrodynamics).  The functional
forms of conservative and friction forces in this DPD models are
dictated by the mesh-free discretization, as well as the input
information of the field theory itself.  We have the impression that
SDPD or its isothermal counterpart MDPD are underappreciated and
underused.  Although these methods are appropriate for fluid systems,
we foresee the use of MDPD many-body potentials of the embedded atom
form also for CG potentials for bonded atom systems.  While CG
potentials depending on the \emph{global} density are potentially a
trap \cite{Louis2002, DAdamo2013}, the inclusion of many-body
functional forms of the embedded atom kind depending on \emph{local}
density is a promissing route to have more transferable CG potentials
\cite{Allen2008}, valid for different thermodynamic points.  This
expectation, though, needs to be substantiated by further research.
In particular, liquid state theory for MDPD may need to be further
developed \cite{Merabia2007, McCarty2014}.

This Perspective  on DPD  also points  to several  open methodological
questions.   

We have already mentioned the open problem of deriving from
microscopic principles the dynamics of Lagrangian fluid particles made
of unbonded atoms.  Once this problem is solved, we will need to face
the next problem of deriving from first principles the coupling of CG
descriptions of bonded and unbonded atoms (a protein in a membrane
surrounded by a solvent, for example).  A derivation from bottom-up of
this kind of coupling in a discrete \emph{Eulerian} setting has been
given recently in \cite{EspanolDonev2015}.

For the simulation of fluids, standard CFD methods equipped with
thermal fluctuations are readily catching up with the mesoscale
\cite{Naji2009, Uma2011, Shang2012, Donev2010, Oliver2013, Donev2014,
  Donev2014a, Plunkett2014, DeCorato2016}.  Methods for coupling
solvents and suspended structures are being devised
\cite{BalboaUsabiaga2014, EspanolDonev2015}, and therefore one may
well ask what is the advantage of a Lagrangian solver based on the
relatively inaccurate SPH discretization over these high quality CFD
methods. Note that CFD methods allow for the rigorous treatment of
limits (incompressibility, inertia-less, \etc) that may imply large
computer savings, and which are difficult to consider in SPH based
methods.  We believe (see Meakin and Xu \cite{Meakin2009} for a
defense of particle methods) that fluid particle models may still
compete in situations where biomolecules and other complex molecular
structures move in solvent environments, because one does not need to
change paradigm: only particles for both, solvent and beads, are used,
with the corresponding simplicity in the codes to be
used. Nevertheless, a fair comparison between Eulerian and Lagrangian
methodologies is still missing.

As SDPD is just SPH plus thermal fluctuations it inherits the
shortcomings of SPH itself.  SPH is still facing some challenges in
both, foundations (boundary conditions) and computational efficiency
\cite{Violeau2016, Zhi-bin2016}. In this respect, a Voronoi fluid
particle model \cite{Serrano2001}, understood as a Lagrangian finite
volume solver may be an interesting possibility both in terms of
computational efficiency and simplicity of implementation of boundary
conditions.  Serrano compared SDPD and a particular implementation in
2D of Voronoi fluid particles \cite{Serrano2006a}.  In terms of
computational efficiency, both methods are comparable because the
extra cost in computing the tessellation is compensated by the small
number of neighbours required, six on average, while in SDPD one needs
20-30 neighbours.

Another interesting area of research is that of multi-scale modeling.
In CFD, one way to reduce the computational burden is to increase the
resolution of the mesh only in those places where strong flow
variations occur, or interesting molecular physics requiring small
scale resolution is taking place.  An early attempt within DPD was
given by Backer \etal\ \cite{Backer2005}.  We envisage that methods
for multi-resolution SDPD will be increasingly used in the future
\cite{Kulkarni2013, Lei2015, Tang2015, Petsev2016}.  Multi-resolution
is a problem of active research also in the SPH community
\cite{Violeau2016}.  Eventually, one would like to hand-shake the
particle method of SDPD with MD as the resolution is decreased
\cite{Petsev2015}.  Note, however, that as the fluid particles become
small (say ``four atoms per particle'') it is expected that the
Markovian property breaks down and one needs to account for
viscoelasticity \cite{Zwanzig1975, Voulgarakis2009b}, either with
additional internal variables \cite{tenBosch1999, Ellero2003,
  Vazquez-Quesada2009pre}, or with ``fictitious particles''
\cite{Davtyan2015}.

Finally, a very interesting research avenue is given by the
thermodynamically consistent (\ie, able to deal with non-isothermal
situations) Mori-Zwanzig EDPD introduced theoretically by Espa\~nol
\etal\ \cite{Espanol2016}. Up to now, CG representations of complex
molecules have only included the location and velocity of the CG beads
or blobs (sometimes its spin  \cite{Li2014}), completely forgeting its
internal energy content. Given the fundamental importance of the
principle of energy conservation, it seems that in order to have
thermodynamically consistent and more transferable potentials, we may
need to start looking at these slightly more complex CG
representations.

\section{Acknowledgments}
We acknowledge A. Donev for useful comments on the manuscript. PE
thanks the Ministerio de Econom\'{\i}a y Competitividad for support
under grant FIS2013-47350-C5-3-R.

\appendix
\section{MDPD consistency}\label{app:mdpd}
In MDPD the potential takes the form described in the main text where
$V(\{\rvec\})=\sum_i\psi(d_i)$, and $d_i=\sum_{j\ne i}W(r_{ij})$.
From this it is easy to show that the forces remain pairwise, with
\begin{equation}
  \Fvec_{ij} = -[\psi'(d_i)+\psi'(d_j)]\,W'(r_{ij})\,\evec_{ij}\,.
  \label{eq:app:fij}
\end{equation}
Note that the weight function here is $W'(r)$.  However, to our
knowledge, there does not exist in the literature a proof of the
\emph{converse}, namely that this relationship between the weight
functions is a \emph{necessary} condition to ensure the existence of
$V(\{\rvec\})$.  We present here such a proof, following the line of
argument in \Refcite{Warren2013}.

We start with a generalised MDPD pairwise force law, with an
(as yet) arbitrary weight function $\wc(r)$,
\begin{equation}
  \Fvec_{ij} = A(d_i,d_j)\,\wc(r_{ij})\,{\hat\rvec}_{ij}\,.
  \label{eq:app:gen}
\end{equation}
We assume the amplitude function $A(d_i,d_j)$ is symmetric since
otherwise $\Fvec_{ij}\ne-\Fvec_{ji}$.  Let us denote partial
derivatives with respect to the first and second density arguments by
$A_{[1,0]}$ and $A_{[0,1]}$.  The symmetry of $A(d_i,d_j)$ then
implies $A_{[1,0]}(d_i,d_j)=A_{[0,1]}(d_j,d_i)$.

A generic radial force law can always be integrated, so we cannot
deduce anything useful just by considering pairs of particles.
Instead, following \Refcite{Warren2013}, let us consider three
isolated, collinear particles, at positions $x_i$ ($i=1\dots3$) such
that $x_1\le x_2\le x_3$.  For this configuration the densities are
$d_1=W(x_{12})+W(x_{13})$, $d_2=W(x_{12})+W(x_{23})$, and
$d_3=W(x_{13})+W(x_{23})$.  The pairwise forces are
$F_{12}=A(d_1,d_2)\,\wc(x_{12})$, $F_{23}=A(d_2,d_3)\,\wc(x_{23})$,
and $F_{13}=A(d_1,d_3)\,\wc(x_{13})$.  Finally, the summed forces on
the particles are $F_1=F_{12}+F_{13}$, $F_2=-F_{12}+F_{23}$, and
$F_3=-F_{13}-F_{23}$.

The existence of a potential implies integrability constraints like
$\partial F_1/\partial x_2-\partial F_2/\partial x_1=0$.  Imposing
these gives rise to an expression which can be simplified (by
consideration of special cases) to a set of requirements for which the
representative case is
\begin{equation}
  \begin{split}
  &\wc(x_{12})\, W'(x_{23})\, A_{[1,0]}(d_1+d_3,d_1)\\
  &\qquad -\wc(x_{23})\, W'(x_{12})\, A_{[1,0]}(d_1+d_3,d_3)=0.
  \end{split}
  \label{eq:app:mix}
\end{equation}
The symmetry relation between $A_{[0,1]}$ and $A_{[1,0]}$ has been
used.  If we are allowed to cancel the $A_{[1,0]}$ functions we are
home and dry, since this implies $\wc(x)\,W'(y) = \wc(y)\,W'(x)$ (for
arbitrary arguments $x$ and $y$), and this can only be true if
$\wc(x)\propto W'(x)$.  However, the $A_{[1,0]}$ functions only cancel
if $A_{[1,0]}(x+y,x)=A_{[1,0]}(x+y,y)$ (for arbitrary arguments $x$,
$y$).  A little thought shows that a sufficient condition for this to
be true is that $A(d_i,d_j)=f(d_i)+f(d_j)$.  This is precisely the
form the force-law takes in \Eqref{eq:app:fij}.  The conclusion is
that in this case $\wc(x)\propto W'(x)$ is a \emph{necessary}
condition for the existence of the many-body potential $V(\{\rvec\})$.
It is also sufficient, since we can absorb the proportionality
constant into the definitions of $d_i$ and $\psi(d)$, and then
explicitly $V(\{\rvec\})=\sum_i\psi(d_i)$.  This proves the claimed
result above.

For another example, we might be tempted to consider
$A(d_i,d_j)=f(d_i+d_j)$, but retaining the weight function
$\wc(x)\propto W'(x)$.  For this choice $A_{[1,0]}(x,y)=f'(x+y)$ and
\Eqref{eq:app:mix} reduces to $f'(2x+y)=f'(x+2y)$.  This is true
for arbitrary $x$ and $y$ if only if $f(x)$ is linear, and therefore the force
law is \defacto\ of the form shown in \Eqref{eq:app:fij}.  Thus, a non-linear
function $f(x)$ would be a bad choice.  For a further
case study, see \Refcite{Warren2013}.

If we fail to satisfy \Eqref{eq:app:mix} then the potential \emph{does
  not exist}.  If the potential does not exist, we lose the
underpinning theory that the stationary probability distribution is
given by \Eqref{gibbs}.  Without this foundation we are in uncharted
waters, and there is no link to established statistical mechanics
and thermodynamics.  

In our opinion, in MDPD the burden rests on the user to display the
$V(\{\rvec\})$ which gives rise to the chosen force law.  The absence
of an explicitly displayed potential leads only to
unwarranted complications.


\begin{thebibliography}{289}
\bibitem[{foo({\natexlab{b}})}]{footnote1}
\bibinfo{note}{Note, however, that the CGAL library is highly developed and
  allows to construct the Voronoi tessellation scaling with the number $N$ of
  nodes as $N\log(N)$.}
\end{thebibliography}

\end{document}